# Self-service Ad-hoc Querying Using Controlled Natural Language*

Janis Barzdins[1], Mikus Grasmanis[1], Edgars Rencis[1], Agris Sostaks[1] and Juris Barzdins[2]

[1]Institute of Mathematics and Computer Science, University of Latvia, Riga, Latvia
{Janis.Barzdins, Mikus.Grasmanis, Edgars.Rencis, Agris.Sostaks}@lumii.lv
[2]Faculty of Medicine, University of Latvia, Riga, Latvia
Juris.Barzdins@lu.lv

**Abstract.** The ad-hoc querying process is slow and error prone due to inability of business experts of accessing data directly without involving IT experts. The problem lies in complexity of means used to query data. We propose a new natural language- and *semistar* ontology-based ad-hoc querying approach which lowers the steep learning curve required to be able to query data. The proposed approach would significantly shorten the time needed to master the ad-hoc querying and to gain the direct access to data by business experts, thus facilitating the decision making process in enterprises, government institutions and other organizations.

**Keywords:** Ad-hoc querying · star ontologies · controlled natural language · hierarchical data

## 1. Introduction and Problem Statement

The amount of data collected by enterprises, government institutions and other organizations grows significantly every year. Data alone do not guarantee a success – data should be transformed into information, and it should be used accordingly in order to succeed. This process has often been referred to as Business Intelligence (BI). Typically, BI tools offer wide possibilities of data analysis, however they require a significant amount of investment and IT expertise. Needless to say that BI processes involve IT experts whose task is to translate business requirements and queries into a language which is understandable by computer. For example, the Children's Clinical University Hospital (Riga, Latvia) collects sensitive data of clinical processes. Data are stored in the relational database and maintained by the team of local IT experts. The IT experts take part in the clinical processes. They translate hospital managers, practitioners and researchers questions into SQL queries. Although there are predefined reports, the business requirements are changing very often and, consequently, IT experts are overloaded. Therefore business decision processes are very slow and error-prone because of miscommunication and hurry. Direct access to data by domain experts would be a solution. However the problem is that domain experts do not possess the required skills to formulate queries by themselves, because of the complexity of query languages used to retrieve answers from data stores. In our previous work [1] we have defined so called "3How" problem which consists of three main problems related to this context:

1) how to describe data to be easily perceived by domain experts;
2) how to query data simply enough for domain experts;
3) how to perform query efficiently enough in order to get an answer to a sufficiently wide class of queries in reasonable time.

Actually, the problem has been relevant for more than 40 years. The SQL (SEQUEL) language [2] is the *de facto* standard of querying relational databases where most of data are being stored at the moment. However, it turned out that the way data are stored and retrieved in the relational databases was too complicated for domain experts. There are similar languages to SQL (e.g. SPARQL for RDF ontologies [3]) which require a very precise formulation of the textual query (both syntax and semantics) and deep knowledge of underlying technology, thus making them too sophisticated to learn and use. Therefore there have been attempts to make wrappers for these languages – e.g. graphical query builders like Graphical Query Designer for SQL Server [4], ViziQuer [5] and Ontology Based

Data Access (OBDA) approach [6], particularly, the OptiqueVQS [7] for SPARQL and RDF databases, or form-based tools using wizards and standard GUI elements (e.g. tables and lists) like SAP Quick Viewer SQVI [8]. There are also other proposals which provide the means for direct data access. One of the most well-known approaches is Self-Service Business Intelligence (SSBI) which was proposed by Microsoft [9]. It provides a rich set of tools (Power BI) allowing the end-user to build sophisticated data visualizations and make data analysis mainly through spreadsheet applications.

Yet, there is a significant drawback of the mentioned approaches, namely, a steep learning curve which is required to learn a new query language and to understand the way data are stored. In order to make the learning of ad-hoc querying easier we propose a new query language which is based on the controlled natural language. We rely on simple *semistar* data ontologies which resemble the structure of documents which are backing up the business processes in the organization. Health records are good examples of such documents in a hospital. *Semistar* ontology defines a vocabulary (terms) which are allowed to use while querying the data. This allows us to control the language in the way which is familiar to the domain expert. Therefore, the complexity of precise understanding of the rich natural language can be reduced in our implementation of natural language query interface. It should be noted that our query system's ability to explain to the user how his query has been understood plays an important role for ad-hoc querying.

Experiments have shown that the proposed query language can be taught in a short time to medical students. Even after a short (2 hours) lecture for a group of medicine students almost every participant could understand the given examples of queries written in the proposed language. Most of the students were also able to formulate queries for the proposed questions about the clinical processes of the hospital. This paper is organized as follows. Section 2 sums up the related work done by other authors. In Section 3 we introduce the concept of *semistar data ontology* which is heavily exploited in designing the proposed query language. The language itself is described in Section 4 where we give its syntax and semantics together with its typical examples. In Section 5 we briefly outline the basics of the implementation of the language, and in Section 6 we describe the practical experiments we have performed to test our language from different points of view. Section 7 concludes the paper.

## 2. Related Work

A viable option to query data simply enough for domain experts (2$^{nd}$ problem of "3How") is a natural language. Therefore in this section we will discuss the existing natural language interfaces to databases (NLIDB-s). A lot of work has been done in this area [10-14]. "Natural language is an effective method of interaction for casual users with a good knowledge of the database, who perform question-answering tasks, in a restricted domain" [15]. It should be noted that the formulation of the precise query itself is a hard problem for users without mathematical background, e.g. Ogden et.al. showed that users would not be able to specify the meaning of "and" clearly enough for unambiguous understanding of a query [16]. Related research [17] in the knowledge base area has reported that there are still lots of problems in the understanding of complex queries.

In order to make an NLIDB system usable by domain experts it is necessary to solve the problem of *linguistic coverage*. People do not know what the computer "knows", i.e. when a natural language is used there is no common context in the conversation between the user and the database [18]. It is very important to explain to users what the database "knows" and what can be asked. Database schemas used by IT experts (like ER models) are too complex and contain too many technical details to be useful for the explanation of the underlying data. Computers, on the other hand, cannot properly understand what users mean by their queries because of richness and ambiguity of the natural language. In order to achieve a *consensus* between the user and the computer an intermediate representation of data schema is needed.

Many NLIDB solutions rely on the user's domain knowledge and meaningful names used in the schema's elements. These solutions search for similarities between the names and the terms written by user [10, 12]. There are NLIDB solutions that use ontologies (*lexicons*, *vocabularies*) to represent the data schema [10, 19, 20]. Ontologies define concepts, their properties and relationships which can be used by user. Ontology is automatically obtained from the relational database schema. Ontology hides some technical details, but still requires the understanding of basic entity-relationship model principles. Thus, the traditional NLIDB approaches have not reached wide usage, at least not for deep querying with nontrivial calculations. Therefore other approaches have to be studied in order to solve the "3How" problem. It is the main goal of this paper.

## 3. Semistar Data Ontologies

Analysis of current situation suggests that it is hopeless to try to develop an easily perceptible query language that can be used on arbitrary ontologies, because such language has not yet appeared after 30 years since the invention of relational databases. Therefore we introduce an important subset of data ontologies called the *Semistar ontologies* (see also [1, 21, 22]). A typical example of a semistar ontology is depicted in Fig. 1. This is a simplified version of data ontology used in Riga Children's Clinical University Hospital (the actually used ontology consists of 25 classes and 142 attributes). Semistar ontologies are data ontologies whose graphical representation corresponds to a star-like structure (having no loops) with a restriction on multiplicities, such that all associations must have the multiplicity equal to 1 at the end of the association that is nearer to the star's center. Semistar ontologies have only one type of associations between basic classes – the "has" relation (e.g. Patient has HospitalEpisodes, HospitalEpisode has TreatmentWards, etc.). Besides basic classes, a semistar ontology can also contain other classes called the classifiers (depicted with white background in Fig. 1). Associations between basic classes and classifier classes are coded as attributes (e.g. familyDoctor: CPhysician).

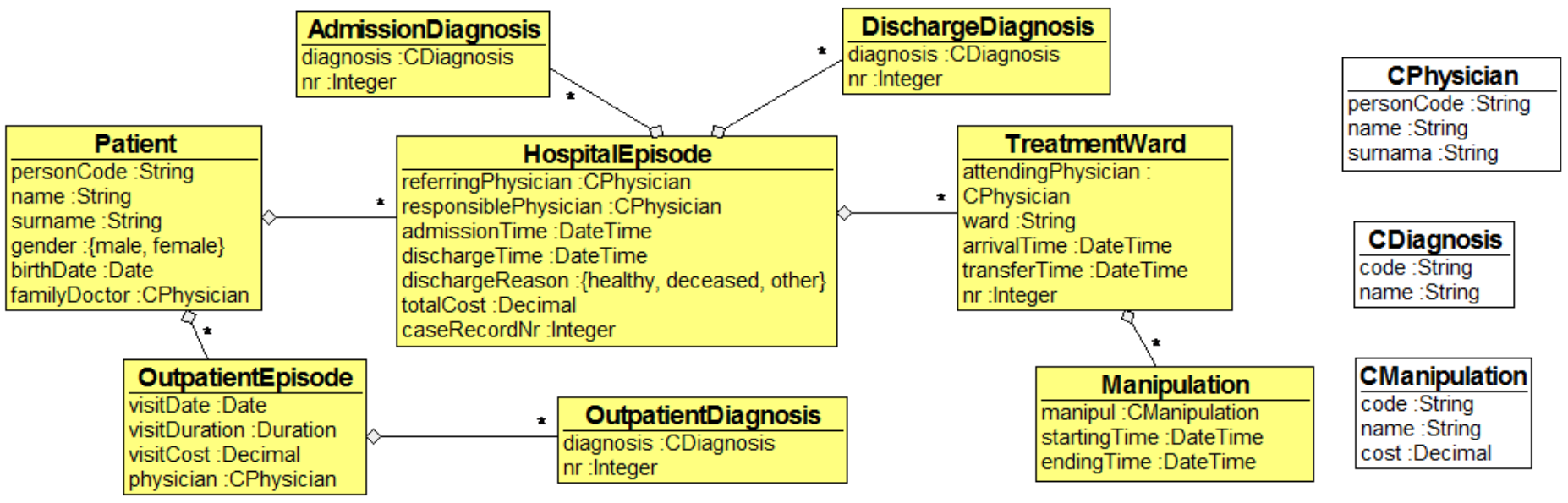

**Fig. 1.** Simplified semistar ontology used in Riga Children's Clinical University Hospital.

Semistar ontology is a practically important and expressive subset of all data ontologies, and practical use-cases often exploit exactly this type of ontologies. As can be seen in Fig. 1, hospital ontology viewed from patients' and physicians' point of view comes out to be a semistar ontology. Our experience shows that even in more general cases, when some ontology is not a semistar ontology, one can usually find an important subset of it to comply to principles of semistar ontology. We can always think of a semistar ontology as a subject-oriented ontology where the role of the subject can be performed by a patient (in case of medical domain), a customer (in case of some service domain), etc.

We allow attributes of basic classes to have two kinds of data types – the primitive types and the classifiers. We use the following predefined data types and operations:

- Integer (e.g. 75, -75), Real (e.g. 0.75, -75.0), operations: +, -, *, /;
- Boolean (true, false), operations: and, or, not;
- String (e.g. "abc"), operations: substring (e.g. "abcde".substring(2,3)="bc");
- Date (e.g. 2015.06.17), unary operations: year(), month(), day(), dayOfWeek(), binary operation: - (e.g. 2015.06.17-2015.05.12 = 1M5D);
- DateTime (e.g. 2015.06.17T10:45), unary operations: year(), month(), day(), hour(), minute(), second(), date(), binary operation: - (subtraction);
- Duration (e.g. 3Y4M5DT6H7M30.25S), unary operations: years(), months(), days(), hours(), minutes(), seconds().

If some attribute *a* has a classifier class as data type and this classifier class has some attribute *k*, then also *a.k* denotes a valid *attribute expression* and its data type will be that of attribute *k*. If *x* is an instance of some class, for which attribute *a* is defined, then also *x.a* (or *x.a.k*, if type of *a* is a classifier class) denotes a valid attribute expression. We can build more complex attribute expressions from simpler ones using the abovementioned operations allowed for the given data types. Some examples of attribute expressions: *personCode, x.personCode, x.familyDoctor.surname, x.admissionTime.month(), (dischargeTime-admissionTime).days(),* etc.

We can now compare two attribute expressions (or constants) to obtain *attribute conditions*, e.g. *personCode=250285-10507, x.personCode=250285-10507, personCode.substring(1,4)=2502, dischargeTime-admissionTime>25d* (meaning – 25 days), *x.birthDate.year()>=1985, familyDoctor<>nil* (a family doctor exists), etc.

## 4. Query Language

### 4.1. Basic Ideas

The query language we propose in this paper is to be used for formulation of ad-hoc queries, meaning queries that can be formulated in one sentence (perhaps together with some subordinate clauses) in natural language, so that the end-user can still understand it very well. The language will work on semistar ontologies. The simplicity of the "has" relation is the main factor, which allows query language to be simple and similar to a natural language. It is therefore convenient to build queries in a controlled natural language. This feature allows the language to be easily perceptible by non-IT specialists.

Let us introduce an example query that will be exploited further in this section – *count Patients, who have at least one HospitalEpisode, which has Manipulation with manipul.code=02078*. This natural language sentence is understandable by the domain expert. Let us now inspect a bit more complicated query: *count Patients, who have at least one HospitalEpisode, which has at least one TreatmentWard, which has at least one Manipulation with manipul.code=02078*. This sentence may cause a certain ambiguity as it is not clear whether the asked *Manipulation* refers to *HospitalEpisode* or to *TreatmentWard*. It could be used in both meanings. In other words, relative pronouns such as "who" and "which" not always give us accurate understanding of what we relate to. To cope with such situations we introduce a concept of so called short name in our controlled natural language. Formally, the short name is a variable over instances of the given class – *count Patients p, where exists p.HospitalEpisode e, where exists e.TreatmentWard t, where exists t.Manipulation m, where m.manipul.code=02078*. Now we are able to specify also the second of abovementioned meanings – *count Patients p, where exists p.HospitalEpisode e, where exists e.TreatmentWard t, where exists e.Manipulation m, where m.manipul.code=02078*. We have also unified other components of the natural language, e.g. we use the keyword "where" instead of "who", "which" and "with", and the keyword "exists" instead of "have/has at least one". The dot notation after the short name must be perceived as the "of" relation – *count Patient p, where exists HospitalEpisode e of Patient p, where exists TreatmentWard t of HospitalEpisode e, where exists Manipulation m of TreatmentWard t, where manipul.code of Manipulation m equals 02078*.

Formally speaking, the short name must be used before every attribute name to get rid of ambiguities. However, in cases when it is clear to which class the particular attribute refers the short name can be omitted. We also allow omitting other features of the language that are not critical for understanding of queries (e.g. one can omit the empty parentheses after the unary Date and DateTime operations like *year()* or *minute()*).

Let us now introduce some basic notations that we will use describe the query language. We will use the terms *parent class* and *child class* to refer to classes that are higher or lower in the "have" hierarchy. For example, the class "TreatmentWard" has two parent classes – "HospitalEpisode" (direct parent) and "Patient" (further ancestor) and one child class "Manipulation". If *x* is an instance of the class "TreatmentWard", then its parent instances will be denoted as *x.HospitalEpisode* and *x.Patient*. In both cases they denote exactly one instance, i.e. that of the class "HospitalEpisode" and of the class "Patient", respectively. We use the same dot notation also for accessing instances of child classes, but in this case we obtain a set of instances. For example, *x.Manipulation* would be a set of manipulations reachable from the given treatment ward *x*.

In more complicated cases another concept of *brother class* is important. If *x* is an instance of the class “HospitalEpisode”, then by *x.HospitalEpisode* we understand *y.HospitalEpisode*, where *y=x.Patient* (i.e. *y* is the closest parent of *x*, which is also parent class of the given class “HospitalEpisode”). Similarly, if *x* is an instance of the class “TreatmentWard”, then *x.OutpatientEpisode=y.OutpatientEpisode*, where *y=x.Patient*.

If *AClass* is an arbitrary class of the ontology, we will use the term *AClass attribute expression* to denote attribute expressions of both *AClass* itself and all of its parent classes (we assume here that parents and children share no common attribute names). We cannot use attribute expressions of child classes here, because there can be many children instances for the given *AClass* instance. We will be able to access these instances by introducing quantors *exists*/*notexists* later.

We can also perceive our query language as an analogue to some many-sorted predicate language with a difference that it is written in such syntax that is more user-friendly. There has been an attempt to create such a language [23], though it has not led to a practical implementation.

### 4.2. Syntax and Semantics of the Query Language

Queries are to be written in a controlled natural language and are based on seven sentence templates. The main part of the templates is the so called *selection condition*, which is a selection condition over instances of the given class. We assume that selection conditions are to be written in a natural language. We describe the used language constructs more formally at the end of this section. However, the sentence templates described in this section can be understood without knowing the precise syntax of selection conditions.

**T1. COUNT AClass [x] WHERE <selection condition>**

Semantics: counts instances of *AClass*, which satisfy the selection condition. Examples:

- *COUNT Patients, WHERE EXISTS HospitalEpisode, WHERE referringPhysician=familyDocto*r (count of patients who have been referred to hospital by their family doctors);
- *COUNT HospitalEpisodes, WHERE dischargeTime-admissionTime>15d* (how many episodes have lasted longer than 15 days);
- *COUNT HospitalEpisodes e1, WHERE EXISTS HospitalEpisode e2, WHERE e1<>e2 AND e2.admissionTime>e1.dischargeTime AND e2.admissionTime-e1.dischargeTime<30d* (how many there have been such episodes, after which the patient has returned to hospital in less than 30 days).

**T2. {SUM/MAX/MIN/AVG/MOST} <attribute expression> FROM AClass [x] WHERE <selection condition>**

Semantics: selects instances of *AClass*, which satisfy the selection condition, calculates the attribute expression for each of these instances obtaining a list to which the specified aggregate function is then applied. Examples:

- *SUM totalCost FROM HospitalEpisodes, WHERE dischargeReason=healthy AND birthDate.year()=2012* (how much successful treatments of patients born in 2012 have cost);
- *MOST diagnosis.code FROM DischargeDiagnoses, WHERE nr=1 AND dischargeReason=deceased* (get the most frequent main (nr=1) death diagnosis).

**T3. SELECT FROM AClass [x] WHERE <selection condition> ATTRIBUTE <attribute expression> ALL DISTINCT VALUES**

Semantics is obvious. Examples:

- *SELECT FROM HospitalEpisodes, WHERE dischargeReason=deceased, ATTRIBUTE responsiblePhysician.surname ALL DISTINCT VALUES*;
- *SELECT FROM DischargeDiagnoses, WHERE nr=1 AND dischargeReason=deceased, ATTRIBUTE diagnosis.code ALL DISTINCT VALUES.*

**T4. SHOW [n/ALL] AClass WHERE <selection condition>**

Semantics: shows n or all instances of *AClass* which satisfy the selection condition.

**T5. `FULLSHOW [n/all] AClass WHERE <selection condition>`**

Semantics: the same as "show", but shows also the child class instances attached to the selected *AClass* instances.

**T6. `SELECT AClass x WHERE <selection condition>, DEFINE TABLE <x-expr'1> [(COLUMN` $C_1$`], …, <x-expr'n> [(COLUMN` $C_n$`)] [, KEEP ROWS WHERE <`$C_i$` selection condition>] [, SORT [ASCENDING/DESCENDING] BY COLUMN` $C_i$`] [, LEAVE [FIRST/LAST] n ROWS]`**

Semantics: selects all instances of *AClass*, which satisfy the selection condition, then makes a table with columns $C_1$ to $C_n$, which for every selected *AClass* instance *x* contains an individual row, which in column $C_1$ contains the value of the <x-expr'1>, …, in column $C_n$ contains the value of the <x-expr'n>. Then it is possible to perform some basic operations with the table like filtering out unnecessary rows, sorting the rows by values of some column and then taking just the first or the last n rows from the table. Examples:

- *SELECT HospitalEpisodes x, WHERE dischargeReason=deceased, DEFINE TABLE x.surname (COLUMN Surname), x.dischargeTime.date() (COLUMN Dying_date), (COUNT x.Manipulation, WHERE manipul.code=02078) (COLLUMN Count_02078), (SUM manipul.cost FROM x.Manipulation, WHERE manipul.code=02078) (COLUMN cost_02078)*;
- *SELECT CPhysicians k, WHERE name=Gatis AND EXISTS HospitalEpisode, WHERE responsiblePhysician=k, DEFINE TABLE surname (COLUMN Physician_surname), (COUNT HospitalEpisodes, WHERE responsiblePhysician=k) (COLUMN Episode_count), (MOST diagnosis.code FROM AdmissionDiagnoses, WHERE nr=1 AND responsiblePhysician=k) (COLUMN Most_frequent_main_diagnosis), KEEP ROWS WHERE Episode_count>5, SORT DESCCENDING BY COLUMN Episode_count, LEAVE FIRST 10 ROWS.*

Let us now talk a bit more precisely about the means for defining columns. The expression *<x-expr'i>* defines the value of column $C_i$ in the row that corresponds to the *AClass* instance *x*. This expression can be defined in one of four ways:

`<x-expr'i> ::= <x-dependent attribute expression> | <x-dependent count expression> | <x-dependent {SUM/MAX/MIN/MOST] expression> | <x-dependent child attribute selector expression>`

`<x-dependent attribute expression>` examples: *x.surname, x.dischargeTime.date().* Prefix "*x.*" can be used before attributes of both *x* and its parents. Semantics is obvious.

`<x-dependent count expression>` examples: *(COUNT x.Manipulations, WHERE manipul.code=02078), (COUNT HospitalEpisodes, WHERE responsiblePhysician=x).* In the first example we use the prefix *x* in "x.Manipulations" to denote that we do not select from the whole set of manipulations, but only from those that are reachable from *x*.

`<x-dependent {SUM/MAX/MIN/MOST} expression>` examples*: (SUM manipul.cost FROM x.Manipulations, WHERE manipul.code=02078), (MOST diagnosis.code FROM AdmissionDiagnoses, WHERE nr=1 AND responsiblePhysician=x).*

`<x-dependent child attribute selector expression>` examples: *(x.DischargeDiagnosis, WHERE nr=1).diagnosis.code, (x.TreatmentWard, WHERE nr=*).ward* (by * we denote the number of the last instance of TreatmentWard connected to the given HospitalEpisode *x*). This is a new kind of construction whose general form is as follows: *(x.<name of x children class A>, WHERE <selection condition>).<name of attribute a of class A>*. Its value is defined in the following way – we start by taking all instances of class A that are reachable from *x*, then select of them those instances that satisfy the selection condition and then create a list of values of the attribute *a* of the selected instances. The most important case here is the one where this list contains only one instance, e.g. in the following table definition example:

*SELECT HospitalEpisodes x, WHERE dischargeReason=deceased, DEFINE TABLE x.surname (COLUMN Surname), x.dischargeTime.date() (COLUMN Dying_date), (x.DischargeDiagnosis, WHERE nr=1).diagnosis.code (COLUMN main_diagnosis), (x.TreatmentWard, WHERE nr=*).ward (COLUMN last_ward).*

**T7.** There are two more cases in the definition of the table, where table rows come from other source, not being instances of some class. Being very similar these two cases form two subtemplates of the last template:

a) **SELECT FROM AClass [a] WHERE <selection condition> ATTRIBUTE <attribute expression> ALL DISTINCT VALUES x, DEFINE TABLE…**
b) **SELECT FROM INTERVAL (start-end) ALL VALUES x, DEFINE TABLE…**

Semantics of both cases is obvious. Examples:

- *SELECT FROM TreatmentWards ATTRIBUTE ward ALL DISTINCT VALUES x, DEFINE TABLE x (COLLUMN Ward), (SUM manipul.cost FROM Manipulations, WHERE ward=x) (COLUMN Cost)*;
- *SELECT FROM INTERVAL (1-12) ALL DISTINCT VALUES x, DEFINE TABLE x (COLUMN Month), (COUNT HospitalEpisodes, WHERE admissionTime.month()=x) (COLUMN Episode_count) (MOST diagnosis.code FROM AdmissionDiagnoses, WHERE nr=1 AND admissionTime.month()=x) (COLUMN Most_frequent_main_diagnosis).*

Let us conclude this section by defining more formally the constructs of a controlled natural language allowed in the selection conditions. They can, of course, be guessed from the examples given above.

<AClass selection expression> ::= AClass [<short name>] [WHERE <selection condition>]

<selection condition> ::= <attribute condition> | <quantor condition> | (<selection condition> {AND|OR} <attribute condition>) | (<selection condition> {AND|OR} <quantor condition>)

<quantor condition> ::= {[NOT]EXISTS | FORALL} [short name.] AClass [short name] [WHERE <selection condition>]

Short name provides a name for the given object and can be any string different from the class and attribute names. The abovementioned grammar provides a formal language (for formulating selection expressions) that is close to a natural language and therefore easily perceptible. From a natural language's point of view selection expressions are sentences in a controlled natural language that exploit both words of a natural language (like *AND*, *OR*, *WHERE*, *EXISTS*, *NOTEXISTS*) and "foreign" words – attribute expressions whose syntax and semantics were described above. The grammar is only needed as a guide how to build the selection expressions. We do not use it in teaching the language to domain experts. We, instead, use the same principle exploited when a child learns to speak a natural language, i.e. learning from examples. It was therefore necessary to first see the sentence templates together with some examples and only afterwards see the formal grammar underlying parts of these templates.

## 5. Implementation of the Query Language

In this section we will lay out the basics of the system underlying the query implementation. The very basic component here is the system metamodel which describes the classes, associations and attributes used in the particular hospital metamodel on which the query is to be executed. A simplified version of the system metamodel can be seen in Fig. 2. This metamodel is coded in Java, where each logical class of the metamodel corresponds to one Java class. There are only two types of multiplicities used for roles in the system metamodel – 1 and *. These are coded as Java attributes of types *A* and *List<A>* respectively (where *A* is the name of the class to which the respective association end is attached).

The overall architecture of the system is depicted in Fig. 3. The system metamodel described above forms the basis of the architecture and serves as an input for three types of generators seen in Fig. 3.

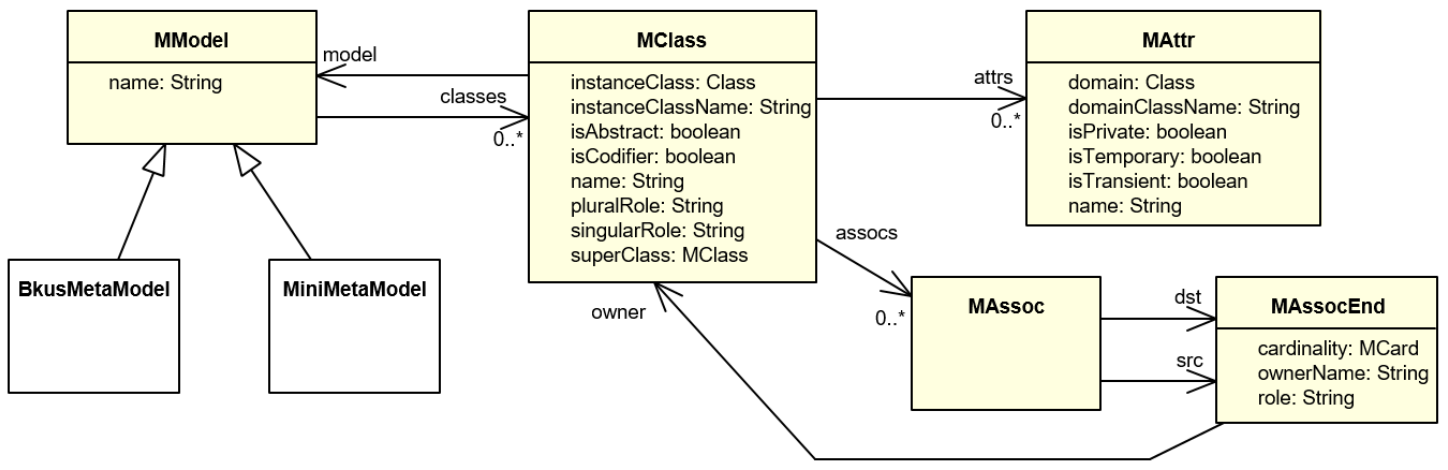

**Fig. 2.** The system metamodel (a simplified version).

The first generator is used for generating information about abstract data types for the querying system. It takes as an input information about classes from the abovementioned system metamodel and gives as an output a set of Java classes together with their respective attributes (including associations). These classes are the ones describing the underlying hospital information system. Let us call them hospital classes. The query will be formed and executed on hospital classes.

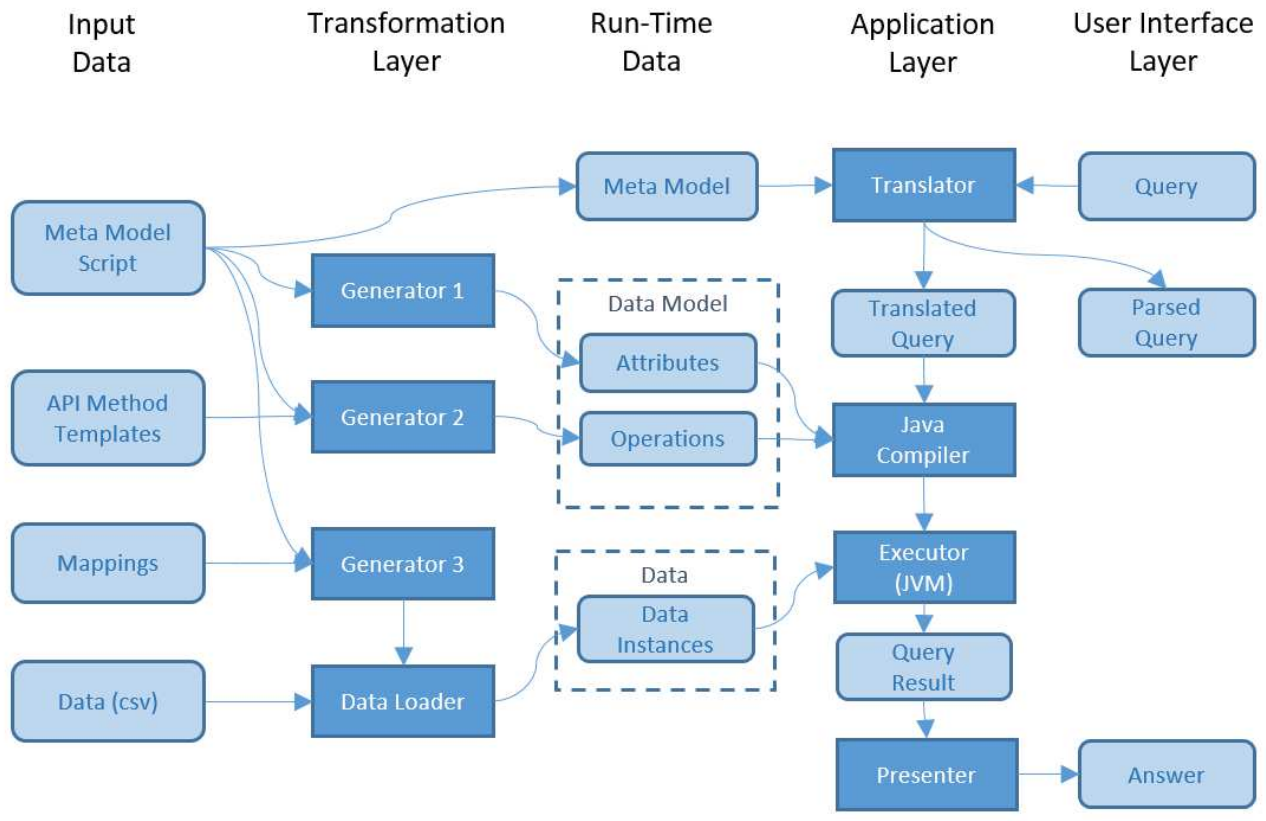

**Fig. 3.** The overall system architecture.

The second generator generates operations for the hospital classes generated by the first generator. It takes as an input the system metamodel and the operation templates and generates necessary operations for hospital classes. These operations lay a foundation to which the natural language queries will be translated. The structure of operation templates and their consisting parts are outside the scope of this paper.

The third generator generates the Java code (called the Loader) that loads particular instances of hospital classes from .csv files into the hospital classes themselves. The generator takes the system metamodel and a mapping from .csv to Java classes as inputs and gives the fully functioning Loader code as an output. The Loader can then read the .csv files obtained from the database of the hospital information system and generate from them respective instances of the hospital classes (each .csv file more or less corresponds to one hospital class). The Loader must be able to cope with incomplete or even incorrect data, which are typical situations, as our experience has shown. For example, the loading process must work well, when some attribute is not set in the .csv file, when the value of some attribute does not comply with the expected data type, when the foreign key points to a non-existent instance of another table etc. The mapping from .csv to Java classes is currently very simple, it only links text fields to attributes. The development of this mapping is a broad topic itself, and we have several ideas about how it can be improved – it can take into account only things needed for the execution of a particular query, it can be bi-directional etc. This is, however, not in the scope of this paper.

All three generators are started one after another, when the system is started. It can take some time, depending on the volume of data to be read from .csv files. When the process has ended all the necessary data are located in RAM, and the system is ready to execute user queries.

Query execution consists of a Read-Evaluate-Print loop – user enters a query in a controlled natural language, it is then translated to Java syntax (largely exploiting the operations generated for hospital classes by the second generator), executed in Java, and finally the obtained result is depicted back to the user. There is one side branch to this general schema as can be seen in Fig. 3 – evaluation of the natural language query is performed by the query translator during the query forming phase, and the parsed query is shown back to the user immediately so that he can alter the query accordingly if necessary. To be able to understand and parse the query, the translator also exploits the system MM to obtain information about class names, attributes and associations.

Large part of the system functionality is composed of the operations generated for hospital classes during the system start-up. Let us now take a bit deeper insight into how these class methods look like. These are functions written in the functional programming style:

- $F<T_1,T_2,\dots,T_n,R>$ – function of n arguments (with types $T_1$, $T_2$, …, $T_n$) whose return type is R;
- pred<T> – predicate of one argument with type T (the same as F<T,bool>).

There are two types of functions, namely, global functions over all instances of some class and more local functions over only those instances of some class that can be reached from some given instance. For example, there is a method *countA(pred<A>)* generated for every class *A*. It returns the total count of instances of class *A*, for which the given predicate returns true. Similarly, there is a method *countAB(b,pred<A>),* which only inspects those instances of class *A* that are reachable through links from the given instance *b* of class *B*. A list of the main methods can be seen in Table 1.

**Table 1.** List of the main methods of hospital classes.

| Prefix of method name | Global method | Method with context |
|---|---|---|
| *count* | + | + |
| *sum* | + | + |
| *min/max* | + | + |
| *avg* | + | + |
| *most* | + | - |
| *countDistinct* | + | - |
| *findOne/findAll* | + | + |
| *any/all* | + | + |
| *map* | + | + |
| *table* | + | + |

In the process of designing the operation structure and syntax two contradicting things were to be taken into account – logical clearness and potential performance. One of the first goals for this system was its ability to respond to queries rapidly (in no more than 1-2 seconds on one year data of Riga Children's Clinical University Hospital) and to do that on a regular computer at hospital, not on a supercomputer. Thus, the implementation efficiency was the main factor that dictated the design of the class methods to which queries are translated. It is therefore not the best solution from the logical clearness point of view.

Concluding the description of the query execution engine, it can be noted that we have developed our own No-SQL embedded in-memory database with a functional query language and with a good API in the form of the generated operations. We have eliminated the need for interpreter and reflection, which has given us an opportunity to improve the performance of query execution. Of course, some improvements have been made possible thanks to the fact that we exploit the data only in read-only mode. We have explored other in-memory databases before and found no solution suitable for our specific needs. We will not analyze deeper other approaches in this paper.

Let us conclude with an example showing the translated result of a query.

Query in our language: *COUNT Patient p, WHERE EXISTS HospitalEpisode e, WHERE EXISTS Manipulation m, WHERE manipul.code=02078*

Translated query in Java: *countPatient(p->anyHospitalEpisodePatient(p,e->anyManipulationHospitalEpisode(e,m->m.manipul.code.equals("02078"))));*

## 6. Proof of Concepts

We showed in Section 3 that semistar ontologies cover quite a wide spectrum of practically important data ontologies including hospital data schemas from patients' and physicians' point of view. In this section we will briefly inspect the other aspects of the "3How" problem mentioned in Introduction, i.e. 1) Is the offered query language expressive enough for practical use-cases and simple enough for understanding by domain experts; and 2) Does the query language have sufficiently efficient implementation?

The first aspect consists of two parts. The expressiveness of the query language was demonstrated by turning it into a working language for Riga Children's Clinical University Hospital when annual reports had to be generated. It turned out to be expressive enough for this task. During the two year period, when it was used for report generation, the language underwent a continuous improvement process. It was important to achieve such a level that managers of wards are able to formulate themselves all the necessary queries without going to a programmer with every $5^{th}$ or $10^{th}$ query to write the desired query in SQL. Results of such queries were either single numbers or data fields, or tables of data fields. In case of tables, we do not undertake all the necessary calculations and table operations provided by other applications such as MS Excel. Our aim is to generate a table containing all the necessary data that can then be exported to a spreadsheet or an R tool (a tool for statistical analysis).

The second part of the first abovementioned aspect regards the possibility for domain experts to learn the language. To test this aspect we performed both individual experiments with potential end-users and group tests. General situation from the language teaching point of view was best demonstrated in an experiment with experienced nurses who study to obtain Master Degree at the Medical Faculty, University of Latvia. We presented a two hour long lecture about the language and the tool for querying the data. One third of that time was devoted to explanation of the underlying data ontology (to explain to non-programmers what is a class, an attribute, etc.). Afterwards the language was explained on examples, and homework was given to test the level of understanding. The homework consisted of two parts. Firstly, students had to understand sentences written in our controlled natural language and to write them in a good really natural language. Secondly, they had to work in the opposite direction turning natural language sentences into our formal language. Results obtained from this experiment can be seen in Table 2. The main conclusion here is that another two hours long lecture after the completion of homework would be beneficial for a better understanding of the proposed query language.

**Table 2.** The results of the experiment

| | Number of students succeeded (n=15) | | | | |
|---|---|---|---|---|---|
| Task execution level (%) | ≥90 | ≥75<90 | ≥50<75 | ≥25<50 | <25 |
| Understanding of queries | 9 | 2 | 2 | 1 | 1 |
| Writing of queries | 3 | 5 | 2 | 3 | 2 |

A very important factor related to the teaching process is the fact that the underlying data ontology was anonymized (from patients', physicians' and wards' point of view), but real. Our experience shows that students being domain experts of this ontology rapidly got very interested in the querying process and started to perceive this as a game. This fact had a beneficial impact on the learning process.

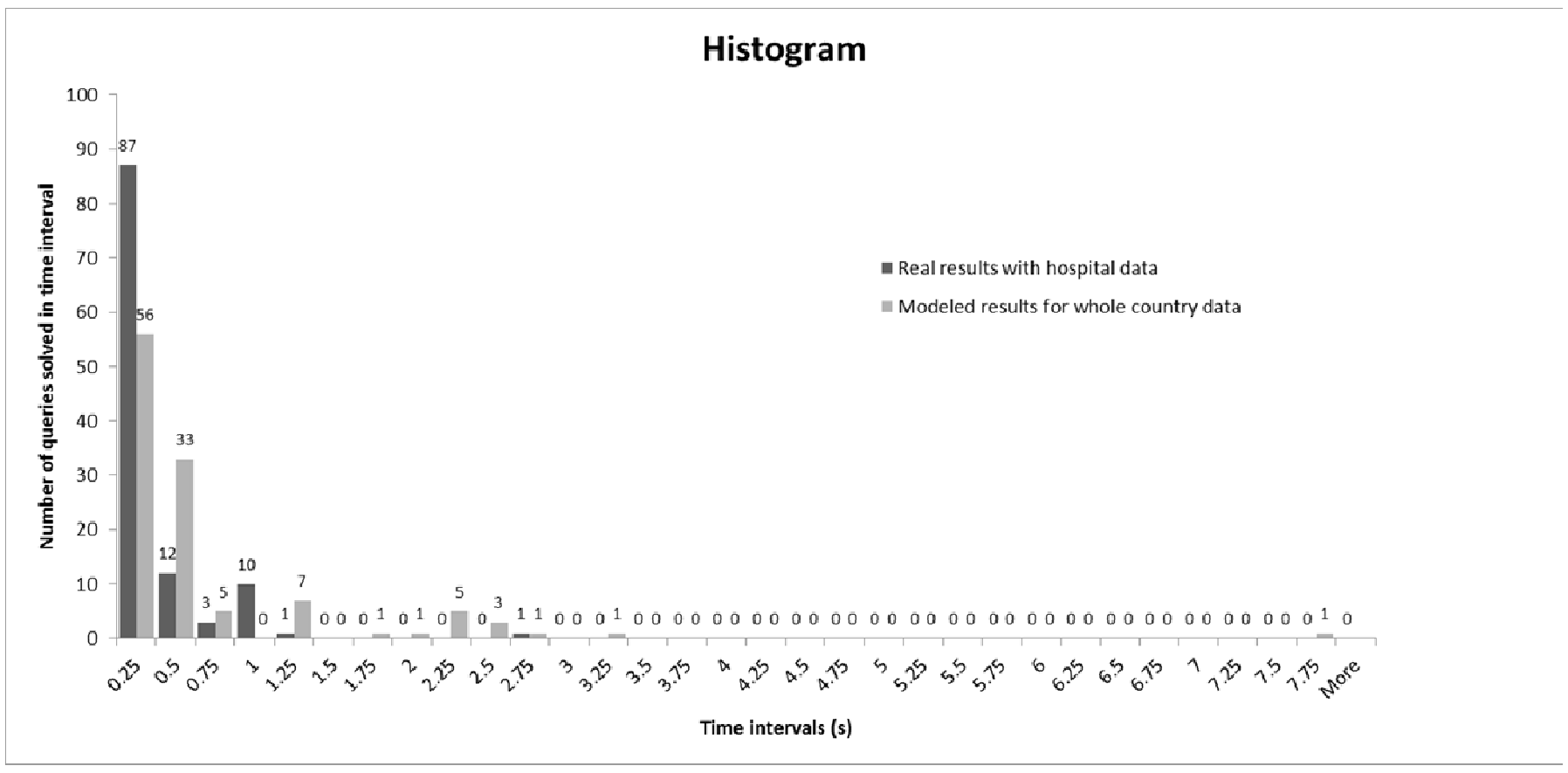

**Fig. 4.** Performance of the query execution.

Finally, let us talk a bit about the ability to implement the language efficiently. The main principles of such implementation were already given in Section 4. To test the performance of our implementation we gathered 120 typical query examples from real annual report analysis of intensive care ward and from discussions with managers of other wards. The complexity of these queries is similar to those demonstrated in Section 4. The volume of data over the period of year 2015 was the following – there were about 35’000 hospital episodes and 70’000 outpatient episodes in Riga Children’s Clinical University Hospital (in total the data took up less than 2GB RAM). The performance on such queries and data volume can be seen in Fig. 4, where queries are sorted in an increasing order by their execution time.

We can see that the vast majority of these 120 queries executes in less than 0.3 seconds. According to statistics there are about 350’000 hospital episodes together in all hospitals in Latvia per year (about ten times more than in Riga Children’s Clinical University Hospital). It means that all these data would take up less than 20 GB RAM. Currently a quad-core computer with 32 GB RAM costs about 1’000 euro. Since the semistar data ontology is granular [21, 22], the query execution can be done in parallel on all four cores thus improving the execution time four times (our experiment seen in Fig. 4 was performed on only one core). We can conclude that the performance of the query execution over data of all the hospitals in Latvia would only be 2.5-3 times slower than it is now providing the ability to answer a vast majority of queries in less than one second.

We are, of course, not limited by only one computer with four cores. We can also use several computers connected via high throughput Ethernet thus reducing the waiting time even more (e.g. one second on ten year data of all Latvian hospitals). Of course, sufficient performance on very large data volumes is another research topic that requires more studies.

By working on the proof of concepts we can conclude that practical testing of our approach has demonstrated that the “3How” problem can be successfully solved at least for the scope of the health system in Latvia.

## 7. Conclusions and Future Work

Practical experiments with our query language have shown that there are yet at least two important features that must be added to increase its usability – subset definition feature (*DEFINE InfectiousDisease = SELECT …*) and attribute definition feature (*DEFINE HospitalEpisode.duration = dischargeTime-admissionTime*). These and similar features are currently under development and require some technical work to be implemented. Another useful feature would be to obtain event distribution in time, which could further be analyzed in MS Excel using its time axis component.

In this paper we have sketched a formal natural language for formulating queries that is still quite a bit controlled. Practical experiments with students have shown that quite a high level of learning is still needed for non-programmers to acquire the skills needed for mastering the language. Our future goals include reducing the level of control of the language and to get closer to the natural language thus allowing end-users formulating queries in much more informal manner. For this to be achieved, strong support for query construction must be developed. Queries would then be formulated very inaccurately (perhaps providing only some basic keywords), and the system could try to understand the query the user has wanted to formulate and offer the resulting query (or more than one potential queries) back to the user for affirmation. This would be the next step towards a really user-friendly query language.

Since data in the medical domain are very sensitive, we are focusing another future research direction towards constructing an advanced access control mechanism. We are planning to exploit the principles of the role-based access control mechanism and apply it to the semistar ontologies where it should also take into account the instance level. We must also consider the decrease in the performance of the query execution by applying different strategies of the access control mechanism and find one that is still satisfactory from the performance point of view.

## Acknowledgments

This work is supported by the Latvian National research program SOPHIS under grant agreement Nr.10-4/VPP-4/11.

Authors are also very thankful to Lolita Zeltkalne for language consulting.